\documentclass[aps,prd,preprint,floatfix,nofootinbib,longbibliography,a4paper]{revtex4-1}
\pdfoutput=1
\usepackage{color}
\usepackage{graphicx}
\usepackage{textcomp}
\usepackage{subfigure}
\usepackage{amsmath,amssymb,array}
\usepackage{enumerate}
\usepackage{hhline}
\newcommand{\mathsym}[1]{{}}

\def\lsim{\:\raisebox{-1.1ex}{$\stackrel{\textstyle<}{\sim}$}\:}
\def\gsim{\:\raisebox{-1.1ex}{$\stackrel{\textstyle>}{\sim}$}\:}

\newcommand{\ba}{\begin{array}}
\newcommand{\ea}{\end{array}}
\newcommand{\be}{\begin{equation}}
\newcommand{\ee}{\end{equation}}
\newcommand{\beqa}{\begin{eqnarray}}
\newcommand{\eeqa}{\end{eqnarray}}

\usepackage{accents}
\usepackage{tikz}
\usepackage{lmodern}
\usetikzlibrary{decorations.pathmorphing}
\tikzset{snake it/.style={decorate, decoration=snake}}
\usepackage{fancybox}
\expandafter\ifx\csname package@font\endcsname\relax\else
\expandafter\expandafter
\expandafter\usepackage
\expandafter\expandafter
\expandafter{\csname package@font\endcsname}%
\fi
\DeclareFontFamily{OT1}{pzc}{}
\DeclareFontShape{OT1}{pzc}{m}{it}%
             {<-> s * [0.900] pzcmi7t}{}
\DeclareMathAlphabet{\mathscr}{OT1}{pzc}%
                                 {m}{it}

\usepackage{epsfig}
\usepackage{graphicx}
\usepackage{amsmath}
\usepackage{amssymb}
\usepackage{epstopdf}
\usepackage{xcolor}
\usepackage{subfigure}
\usepackage{bm}
\usepackage{dsfont}
\usepackage{hyperref}
\usepackage{braket}
\usepackage{todonotes}
\usepackage{comment}
\usepackage{tikz}
\usepackage{pgfplots}
\pgfplotsset{compat=newest}
\newcommand{\bea}{\begin{eqnarray}}
\newcommand{\eea}{\end{eqnarray}}

\def\eq#1{{Eq.~(\ref{#1})}}

\begin{document} 

\title{Signature of neutrino mass hierarchy in gravitational lensing}

\author{Himanshu Swami}
\email{himanshuswami@iisermohali.ac.in}
 
\author{Kinjalk Lochan}
\email{kinjalk@iisermohali.ac.in}
\affiliation{Department of Physical Sciences, Indian Institute of Science Education \& Research (IISER) Mohali, Sector 81 SAS Nagar, Manauli PO 140306 Punjab India.}

\author{Ketan M. Patel}
\email{kmpatel@prl.res.in}
\affiliation{Physical Research Laboratory, Navarangpura, Ahmedabad-380 009, India.\\}

\bigskip
\begin{abstract}
In flat spacetime, the vacuum neutrino flavour oscillations are known to be sensitive only to the difference between the squared masses, and not to the individual masses, of neutrinos. In this work, we show that the lensing of neutrinos induced by a gravitational source substantially modifies this standard picture and it gives rise to a novel contribution through which the oscillation probabilities also depend on the individual neutrino masses. A gravitating mass located between a source and a detector deflects the neutrinos in their journey, and at a detection point, neutrinos arriving through different paths can lead to the phenomenon of interference. The flavour transition probabilities computed in the presence of such interference depend on the individual masses of neutrinos whenever there is a non-zero path difference between the interfering neutrinos. We demonstrate this explicitly by considering an example of weak lensing induced by a Schwarzschild mass. Through the simplest two flavour case, we show that the oscillation probability in the presence of lensing is sensitive to the sign of $\Delta m^2 = m_2^2 -m_1^2$, for non-maximal mixing between  two neutrinos, unlike in the case of standard vacuum oscillation in flat spacetime. Further, the probability itself oscillates with respect to the path difference and the frequency of such oscillations depends on the absolute mass scale $m_1$ or $m_2$. We also give results for realistic three flavour case and discuss various implications of gravitationally modified neutrino oscillations and means of observing them.
\end{abstract} 

\maketitle

\section{Introduction}
\label{sec:intro}
Neutrino oscillation phenomena has provided the most useful platform to study the fundamental properties of neutrinos. The analysis of neutrino oscillation data collected from the  solar, atmospheric and reactor neutrinos (see for example \cite{Capozzi:2013csa,deSalas:2017kay,Esteban:2018azc}) have established that (a) there exist at least three flavours of weakly interacting neutrinos, (b) neutrinos are massive, and (c) their mass eigenstates are different from their flavour eigenstates. However, in this process, we also learn that the neutrino oscillation probabilities  depend only on the difference of the square of neutrino masses and not on their individual masses. Therefore, one cannot infer the absolute neutrino mass scale from the oscillation experiments. Further, the current experiments \cite{PhysRevD.98.030001} based on oscillations have measured $\Delta m_{21}^2$ (where $\Delta m_{ij}^2 \equiv m_i^2-m_j^2$) and $|\Delta m_{31}^2|$  which leaves two  possibilities: either $m_1<m_2<m_3$ or $m_3<m_1<m_2$ known as normal or inverted ordering, respectively,  where $\mathit{ m_1, m_2, m_3}$ denote masses corresponding to the neutrino mass eigenstates.

The non-oscillation experiments, like measuring end-point energy of electron in the nuclear beta decay \cite{Aker:2019uuj} or measuring the rate of neutrino-less double beta decay (only if neutrinos are Majorana fermions) \cite{Dolinski:2019nrj}, can provide direct evidence for the neutrino mass scale. Moreover, the cosmological observations provide a constraint on the sum of all three neutrino masses \cite{Lesgourgues:2014zoa}. Currently, all these experiments lead only to an upper bound on the mass of the lightest neutrino. The strongest cosmological constraints \cite{Vagnozzi:2017ovm,Choudhury:2018byy} imply the lightest neutrino mass $\lsim 0.05$ eV while the latest results from beta-decay experiment \cite{Aker:2019uuj} lead to a much weaker limit, $< 1$ eV.

Theoretical studies of neutrino oscillations are performed and the corresponding experimental data are interpreted mostly in the regime of flat spacetime. The gravitational effects on neutrino propagation have been explored theoretically in somewhat details \cite{Ahluwalia:1996ev,Ahluwalia:1996wb,Grossman:1996eh,Bhattacharya:1996xb,Luongo:2011zza,Geralico:2012zt,Koutsoumbas:2019fkn}. The phenomenological implications of gravitational potential on the neutrino propagation along geodesics are discussed in \cite{Cardall:1996cd,Fornengo:1996ef}. It was shown that the gravitational redshift increases the effective oscillation length of neutrinos. Effects like spin-flip or helicity transitions \cite{Sorge:2007zza,Lambiase:2005gt}, flavour oscillations in case of two and three flavours \cite{Zhang:2016deq} and possible violation of equivalence principle \cite{Lambiase:2001jr,Bhattacharya:1999na} have also been investigated. One of the interesting effects is the lensing of neutrino oscillation probabilities due to gravity. In this case, different trajectories of neutrinos  around a massive astrophysical body are focused on a common point where neutrino flavour oscillation probability is computed. This is first studied in the context of Schwarzschild geometry in \cite{Fornengo:1996ef} and further details are explored in \cite{Crocker:2003cw,Alexandre:2018crg,Dvornikov:2019fhi}.

We explore gravitational lensing of neutrinos for its potential to reveal the absolute neutrino mass scale. The lensing is studied in the context of Schwarzschild geometry in which neutrinos in their travel to the point of observation from a source adopt different trajectories around a gravitating source and get lensed at a common point arriving with different path lengths and hence different phases. The resulting path difference between the accumulating neutrinos results in interference of oscillation probabilities at the point of observation, and it depends not only on the difference of the squared masses but also on the absolute masses of the neutrinos, in general. The qualitative and quantitative aspects of these interference effects are studied in the context of simplified two flavour oscillation case.  We show that the effects of this path dependency seeps into the normalization of the wavefunction as well and hence the overall probability not only cares about the individual masses but the path information as well. We develop the observables which reflect these dependencies and discuss the methodology for obtaining individual mass information from them. We also give results for the three flavour case.

The outline of the paper is as the following. We briefly review the neutrino oscillations in flat and curved spacetime in section \ref{sec:review}. The gravitational lensing effects on neutrino oscillations are discussed in section \ref{sec:lensing}. In section \ref{sec:2FAnalytic}, we discuss explicitly two flavour oscillation case and its qualitative and quantitative features. The results for three flavour case are discussed in section \ref{sec:3FNum}. The study is concluded in section \ref{sec:concl} with a discussion.

\section{Neutrino oscillations in flat and curved spacetime}
\label{sec:review}
In weak interactions, neutrinos are produced and detected in flavour eigenstates denoted by $|\nu_\alpha\rangle$, where $\alpha=e,\mu,\tau$. The flavour and mass eigenstates $|\nu_i\rangle$, with $i=1,2,3$, are related by
\be \label{sup}
|\nu_\alpha\rangle = \sum_{i} U^*_{\alpha i}\, |\nu_i\rangle\,,\ee
where $U$ is $3\times 3$ unitary matrix. In diagonal basis of the charged leptons, $U$ is identified with the leptonic mixing matrix \cite{Esteban:2018azc}. Assuming neutrino wave-function as a plane wave, its propagation from source $S$ to detector $D$, located at spacetime coordinates $(t_S,{\bf x}_S)$ and $(t_D,{\bf x}_D)$ respectively, is described by
\be \label{evolution}
|\nu_i (t_D,{\bf x}_D)\rangle = \exp(-i \Phi_i)\,|\nu_i (t_S,{\bf x}_S)\rangle\,.
\ee
If neutrinos are produced initially in the flavour eigenstate $|\nu_\alpha\rangle$ at $S$, then after travelling to $D$ the probability of the change in neutrino flavour from $\nu_\alpha \to \nu_\beta$ at the detection point is given by
\be \label{Prob}
{\cal P}_{\alpha \beta} \equiv \left|\langle \nu_\beta| \nu_\alpha (t_D,{\bf x}_D)\rangle \right|^2 = \sum_{i,j}U_{\beta i} U^*_{\beta j} U_{\alpha j} U^*_{\alpha i}\, \exp(-i(\Phi_i - \Phi_j))\,.
\ee
The change in flavour can occur if $\Phi_i \neq \Phi_j$. Different neutrino mass eigenstates develop different phases $\Phi_i$ because of difference in their mass and energy/momentum which ultimately gives rise to neutrino oscillation phenomena \cite{Akhmedov:2009rb}.

In flat spacetime, the phase is given by
\be \label{flatPHI}
\Phi_i= E_i(t_D-t_S) - {\bf p}_i\cdot({\bf x}_D-{\bf x}_S)\,.
\ee
It is typically assumed that all the mass eigenstates in a flavour eigenstate initially produced at the source have equal momentum or energy \cite{Akhmedov:2009rb,Akhmedov:2010ua}. Either of these assumptions together with $(t_D-t_S) \simeq |{\bf x}_D-{\bf x}_S|$ for relativistic neutrinos ($E_i \gg m_i$) leads to 
\be \label{flatPHI2}
\Delta \Phi_{ij} \equiv \Phi_i-\Phi_j \simeq \frac{\Delta m_{ij}^2}{2 E_0}\, |{\bf x}_D-{\bf x}_S|,
\ee
where $\Delta m_{ij}^2 \equiv m_i^2 - m_j^2$. $E_0$ is the average energy of the relativistic neutrinos produced at $S$. The oscillation probability ${\cal P}_{\alpha \beta}$ therefore depends on the difference of squared masses and not on the absolute masses of the neutrinos in this case. In other words, a universal shift in the squared masses by a constant, i.e. $m_i^2 \to m_i^2 + C$, leaves the expression of ${\cal P}_{\alpha \beta}$ unchanged. Substitution of Eq. (\ref{flatPHI2}) in Eq. (\ref{Prob}) and further simplification considering only two flavours of neutrinos lead to the following well-known oscillation formula:
\be \label{}
{\cal P}_{e\mu} = \sin^2 2\alpha\, \sin^2\left( \frac{\Delta m_{12}^2\, L}{4 E_0} \right)\,
\ee
where $L=|{\bf x}_D-{\bf x}_S|$. The angle $\alpha$ parametrizes the $2 \times 2$ matrix 
\bea
U \equiv
  \begin{bmatrix}
    \cos{\alpha} & \sin{\alpha}  \\
   -\sin{\alpha}&\cos{\alpha}  
  \end{bmatrix},
\eea
relating the flavour and mass eigenbases for this case.

Modification in neutrino propagation caused by curvature of spacetime has been discussed in \cite{Cardall:1996cd}. In a curved spacetime, the expression of phase in Eq. (\ref{flatPHI}) can be replaced by its covariant form
\be \label{def:phase}
\Phi_i = \int_S^D\, p_\mu^{(i)}\, dx^\mu\,, \ee
where $p_\mu^{(i)}$ is the canonical conjugate momentum to the coordinate $x^\mu$ for the $i^{\rm th}$ neutrino mass eigenstate and it is given by
\be \label{def:p}
p_\mu^{(i)} = m_i\, g_{\mu \nu} \frac{dx^\mu}{ds}\,. \ee
Here, $g_{\mu \nu}$ is the metric tensor and $ds$ is the line element along the neutrino trajectory. Neutrino oscillation probability can be obtained by evaluating the phase $\Phi_i$ for given gravitational field and neutrino trajectories and substituting it in Eq. (\ref{Prob}). For example, such studies have been performed in case of a gravitational field of a static spherically symmetric object described by the Schwarzschild metric \cite{Cardall:1996cd,Fornengo:1996ef}. The line element in this case is given by
\be \label{sczld}
ds^2=B(r)\, dt^2 -\frac{1}{B(r)}\, dr^2 - r^2 d\theta^2 -r^2\, \sin^2\theta\, d\phi^2\,,
\ee
where $B(r) = 1-2GM/r \equiv 1 -R_s/r$, and $G$ and $M$ are Newtonian constant and mass of the gravitating object, respectively and $R_s$  its Schwarzschild radius. 

For simplicity, the motion of neutrinos can be chosen confined on $\theta= \pi/2$ plane as the gravitational field in this case is isotropic. Then the oscillation phase developed by $j^{\rm th}$ neutrino mass eigenstate, $\nu_j$, while travelling from the source $S(t_S,r_S,\phi_S)$ to detector $D(t_D,r_D,\phi_D)$, can be estimated using
\be \label{}
\Phi_j = \int_S^D\left( E_j(r)\, dt - p_j(r)\, dr - J_j(r)\, d\phi \right)\,,
\ee
where $E_j(r) \equiv p_t^{(j)}$, $p_j(r) \equiv - p^{(j)}_r$ and $J_j(r) \equiv - p_\phi^{(j)}$. In this calculation of phase, the classical trajectory from the source to the detector is employed. Being a quantum analysis such usage of ``classical" trajectories seems unjustified and there has been a debate about it \cite{Cardall:1996cd, Lipkin:2000mz} in the literature. However, as we show in Appendix \ref{App:Eikonal}, this approximation can be justified for a relativistic quantum particle in the regime of sufficiently weak gravitational field. This also goes on to suggest that standard treatment of neutrino flavour oscillations can not be applied for strong lensing scenarios, where most of the interesting physics happens around the photon sphere ($R_{ph} = 3 R_s/2)$. Further, being massive, neutrinos do not really travel along the null rays and if one correctly employs the mass effects, we get an extra factor of 2 in the oscillation phase (as also  reported 
in \cite{Bhattacharya:1999na, Chakraborty:2015vla})\footnote{This remains true in the flat spacetime too.}. In order to keep tune with bulk of the existing literature on neutrino oscillation, we will employ the quantum mechanical treatment for weak lensing study with null ray approximation. However, the effects of lensing that we are going to discuss in this work remain qualitatively unchanged if one adopts massive trajectories. 

The phase $\Phi_j$ is explicitly computed for neutrinos traveling along the light-ray trajectory in \cite{Fornengo:1996ef}. For radial propagation, one obtains 
\be \label{Phi_radial} 
\Phi_j \simeq \frac{m_j^2}{2 E_0}\, |r_D - r_S|\,
\ee
where $|r_D - r_S|$ is coordinate difference and it is different from the physical distance in non-flat spacetime. To evaluate the phase for general light-ray trajectory,  it is convenient to write the angular momentum $J_j$ in terms of the energy $E_j$, asymptotic velocity $v_j$ of the corresponding neutrino and the impact parameter $b$ (the shortest distance of the undeflected trajectory). Assuming $GM \ll r$ (weak gravity limit) and $b \ll r_{S,D}$, one finds \cite{Fornengo:1996ef}
\be \label{Phi_non-radial}
\Phi_j \simeq \frac{m_j^2}{2 E_0} (r_S + r_D) \left(1- \frac{b^2}{2 r_S r_D} + \frac{2 GM}{r_S + r_D} \right)\,. \ee
It is easy to see that the phase difference $\Delta \Phi_{jk}$ obtained from Eq. (\ref{Phi_radial}) or (\ref{Phi_non-radial}) depends on $\Delta m^2_{jk}$. Therefore, the oscillation probability even in curved spacetime is also invariant under the shift $m_i^2 \to m_i^2 + C$, similar to the case in flat spacetime discussed earlier. However, as we discussed previously, in curved spacetime, there is a possibility of gravitational lensing as well, which brings in path difference too into play, which leads to breaking of this shift symmetry, as we see in the next section.

\section{Gravitational lensing of neutrinos in Schwarzschild spacetime}
\label{sec:lensing}
\begin{figure}[!ht]
    \begin{center}
        \begin{tikzpicture}[scale=1.5]
            \draw[-] (-5,0) -- (5,0);
            \draw[-] (-5,0) --  (237/73,225/73);
          \draw[-] (-45/73, 120/73)--(4,1);
           \draw[dashed,->] (0,0) -- (-1.5, 4);
           \draw[black,bend left=15] (-5,0) to (4,1);
           \draw[dashed,->] (0,0) to (4,3/2);
           \draw[-] (0,0) -- (0,15/8);
           \draw[dotted,.](4,1) -- (237/73,225/73);
             \draw[-] (0,0) -- (4,1);
           
           \node[label=below: Schwarzschild mass](4) at (0,0) {};
            \node[label=below: Source ](5) at (-5,0) {};
             \node[label=below: Observer](6) at (4,1) {};
              \node[label=above:$\varphi$](7) at (-4,-0.15) {};
               \node[label=right:$\delta$](8) at (0.1-45/73, 0.05+120/73) {};
                  \node[label=above:$y'$](9) at (-1.5,4) {};
                   \node[label=right:$x'$](10) at (4,1.5) {};
                   \node[label=above:$y$](11) at (0,15/8) {};
                   \node[label=right:$x$](12) at (5,0) {};
                    \node[label=left:$b$](13) at (-45/146,60/73) {};
                     \node[label=right:$O$](14) at (4,1) {};
                       \node[label=above:$\varphi$](15) at (-0.1,0.5) {};
                        \node[label=below:$r_S$](15) at (-02.5,0) {};
                         \node[label=below:$r_D$](15) at (2,0.5) {};
        \end{tikzpicture}
    \end{center}
    \caption{Diagrammatic representation of weak lensing of neutrinos.}
    \label{fig:lensing01}
\end{figure}
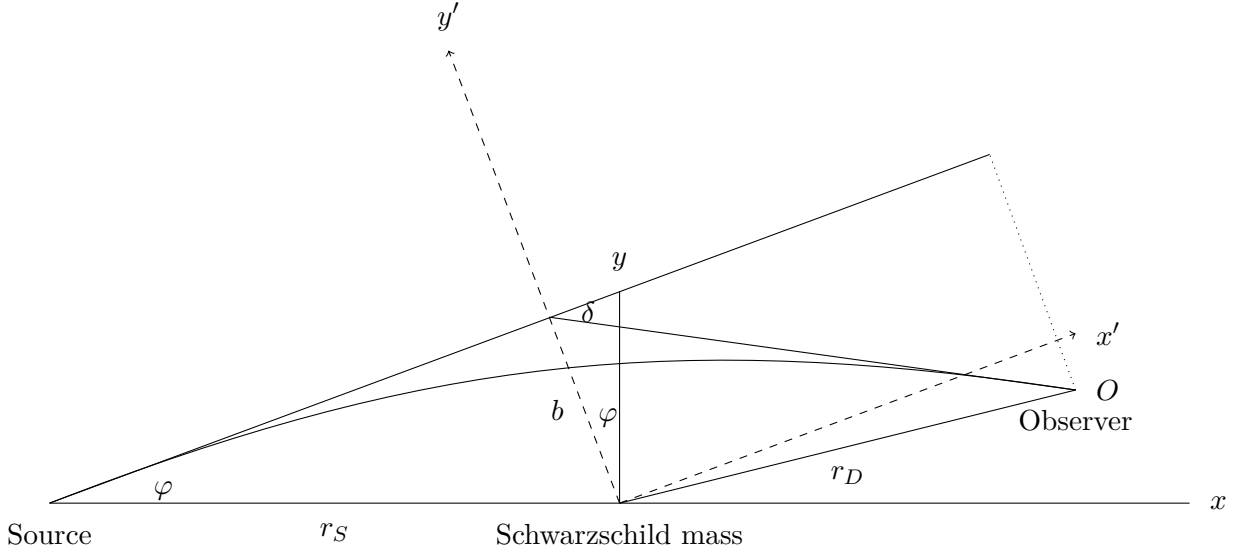

For neutrinos propagating non-radially around the gravitating source, the dependence of the phase on the impact parameter $b$ gives rise to novel effects when the lensing occurs. Let us consider a Schwarzschild black hole as gravitational lens situated between the neutrino source and detector as depicted in Fig. \ref{fig:lensing01}.
In curved spacetime,  neutrinos of a given mass eigenstate $\nu_i$ may travel through different classical paths and meet at the common detection point $D$. The neutrino flavour eigenstate, propagated from the source to detector through different paths denoted by $p$, is given by
\be \label{mod_evolution}
|\nu_\alpha (t_D,{\bf x}_D)\rangle = N  \sum_{i} U_{\alpha i}^* \sum_p \exp(-i \Phi_i^p)\,|\nu_i (t_S,{\bf x}_S)\rangle\,,\ee
where $\Phi_j^p$ denotes the same phase as given in Eq. (\ref{Phi_non-radial}) but contains the path dependent parameter $b_p$. If the neutrino is produced in $\alpha$ flavour eigenstate at the source $S$ then the probability of it being detected in $\beta$ flavour at the detector location, is given by
\be \label{P_lensing}
{\cal P}^{\rm lens}_{\alpha \beta} = |\langle \nu_\beta| \nu_{\alpha}(t_D,{\bf x}_D)\rangle|^2 = |N|^2 
\sum_{i,j} U_{\beta i} U^*_{\beta j} U_{\alpha j} U^*_{\alpha i}\, \sum_{p,q}  \exp\left(-i \Delta \Phi^{pq}_{ij}\right)\,,
\ee
where, 
\be \label{N}
|N|^2 = \left( \sum_i |U_{\alpha i}|^2 \sum_{p,q}\exp\left(-i \Delta \Phi^{pq}_{ii}\right)\right)^{-1}\,.
\ee

The phase difference can be conveniently parametrized, using Eq. (\ref{Phi_non-radial}) in two parts which depend on either the mass difference $\Delta m^2_{ij}$ or the path difference $\Delta b_{pq}^2$, as
\be \label{Delta_Phi}
\Delta\Phi_{ij}^{pq} = \Phi_i^p-\Phi_j^q = \Delta m^2_{ij}\, A_{pq}\,+\,\Delta b_{pq}^2\, B_{ij}\,, \ee
where
\beqa \label{}
A_{pq} & = & \frac{(r_S+r_D)}{2 E_0} \left( 1 + \frac{2 GM}{r_S + r_D} - \frac{\sum b_{pq}^2}{4 r_S r_D} \right)\,, \nonumber \\
B_{ij} & = & - \frac{\sum m_{ij}^2}{8 E_0} \left( \frac{1}{r_S} + \frac{1}{r_D}\right)\,.
\eeqa
Here, $\sum b_{pq}^2 = b_p^2 + b_q^2$ and $\sum m_{ij}^2 = m_i^2 + m_j^2$. Clearly, by construction, the parameters $A_{pq}$ and $B_{ij}$ are symmetric under the interchange of their respective indices $A_{pq} = A_{qp}; B_{ij} = B_{ji}$. Further, the definition implies $\Delta\Phi_{ii}^{pq} = \Delta b_{pq}^2 B_{ii}$,  $\Delta\Phi_{ij}^{pp} = \Delta m_{ij}^2 A_{pp}$ and $\Delta\Phi_{ij}^{pq} = -\Delta\Phi_{ji}^{qp}$.

It can be seen from Eq. (\ref{Delta_Phi}) that the oscillation probability expression (given in Eq. (\ref{P_lensing}) as well) depends on $\sum m_{ij}^2$ through path difference $\Delta b_{pq}^2$. Therefore, it is evident that for a choice of location of detector, for which $\Delta b_{pq}^2$ vanishes, the oscillation probability ${\cal P}^{\rm lens}_{\alpha \beta}$ is invariant under the shift $m_i^2 \to m_i^2 + C$. The locations, for which $\Delta b_{pq}^2 \neq 0$, break this invariance, and the shift implies $B_{ij} \to B_{ij} + 2C$ in these cases. A generic non-collinear configuration of the source, the Schwarzschild mass (lens) and the point of detection, is therefore, expected to retain the information of $\sum m_{ij}^2$ as well. Therefore, we evaluate this dependence of the oscillation probability at a generic point in the source-lens-detector plane, in the weak field limit and obtain observable aspects of it. Substitution of Eq. (\ref{Delta_Phi}) into (\ref{P_lensing}) leads to
\be \label{P_lensing_1}
{\cal P}^{\rm lens}_{\alpha \beta} = \frac{\sum_{i,j} U_{\beta i} U^*_{\beta j} U_{\alpha j} U^*_{\alpha i} \Bigg( \sum_{p}  \exp\left(-i \Delta m^2_{ij}A_{pp} \right) + 2\sum_{p> q}\cos\left( \Delta b_{pq}^2\, B_{ij}\right)\exp\left(-i \Delta m^2_{ij}\, A_{pq}\right)\Bigg)}{N_{\rm path} + \sum_i |U_{\alpha i}|^2 \sum_{q>p} 2 \cos(\Delta b_{pq}^2 B_{ii})}\,.
\ee
The above expression leads to conservation of total probability, i.e. $\sum_{\beta}{\cal P}^{\rm lens}_{\alpha \beta} = 1$. For
simplicity, we consider neutrino propagation in $\theta=\pi/2$ plane. In this case, $N_{\rm path} = 2$ and
the general expression of ${\cal P}^{\rm lens}_{\alpha \beta} $ for$N$ flavour reduces to
\beqa \label{P_lensing_2}
{\cal P}^{\rm lens}_{\alpha \beta} &=& |N|^2 \Big( 2 \sum_{i} |U_{\beta i}|^2 |U_{\alpha i}|^2(1+\cos(\Delta b^2 B_{ii})) + \sum_{i,j\neq i}  U_{\beta i} U^*_{\beta j} U_{\alpha j} U^*_{\alpha i} \Big.\, \nonumber \\
& \times &  \Big.  \left(\exp(-i \Delta m_{ij}^2 A_{11})+\exp(-i \Delta m_{ij}^2 A_{22})+2\cos( \Delta b^2 B_{ij})\, \exp(-i \Delta m_{ij}^2 A_{12}) \right) \Big)\,,
\eeqa
with 
\be 
|N|^2 =\left(2 +  2\sum_i |U_{\alpha i}|^2 \cos(\Delta b^2 B_{ii})\right)^{-1}\, \label{Normalization}
\ee
where $\Delta b_{12}^2 \equiv \Delta b^2$. 

Many apparently similar versions of Eq. (\ref{P_lensing_2}) are available in literature \cite{Fornengo:1996ef, Crocker:2003cw, Alexandre:2018crg}, however with some subtle differences.  The role of normalization in Eq. (\ref{Normalization}) has been neglected in \cite{Fornengo:1996ef,Alexandre:2018crg} which is very important in understanding the neutrino oscillation interference effects as we show in the following sections. From Eq. (\ref{P_lensing_2}), it is easy to verify that  with the proper normalization,  transition and survival probabilities sum up to unity. Since the normalization depends on $\Delta b^2$, one expects it to be path dependent in general. The normalization in Eq. (\ref{Normalization}) also depends on the neutrino mixing parameters unlike the one obtained earlier in \cite{Crocker:2003cw}. The expression of ${\cal P}^{\rm lens}_{\alpha \beta}$  in Eq. (\ref{P_lensing_2}) also differs from that obtained in \cite{Crocker:2003cw} where it is  assumed to be factorizable into two parts: one which depends only on the neutrino path difference while the other depends solely on the neutrino mass difference. We find that such a factorization is not possible in general. Further, unlike in \cite{Alexandre:2018crg}, the expression in Eq. (\ref{P_lensing_2}) also captures the oscillation profile in the pre-lensing phase through its dependency on source location $r_S$. This is a crucial difference, since taking the flat space limit $M \rightarrow 0$ does not appropriately bring out the dependence on the distance between the source and detector in \cite{Alexandre:2018crg} and misses out the phase information during its journey from the source to the lens part. Therefore, we can proceed with Eq. (\ref{P_lensing_2}) to study the dependency of transition on the individual neutrino masses.

\section{Absolute neutrino mass dependent effects in lensing: Two flavour case}
\label{sec:2FAnalytic}
We now discuss the simplest case of two neutrino flavours in order to understand the qualitative difference that arise through lensing effects. In this case, the probability for $\nu_e \to \nu_\mu$ transition obtained from the general expression Eq. (\ref{P_lensing_2}) is as the following.
\beqa \label{P_lensing_3}
{\cal P}^{\rm lens}_{e \mu} &=& |N|^2 \sin^2 2\alpha\, \Big( \sin^2\left(\Delta m^2 \frac{ A_{11}}{2} \right) + \sin^2\left(\Delta m^2 \frac{A_{22}}{2} \right) \Big.\, \nonumber \\
& - &  \Big. \cos(\Delta b^2 B_{12})\, \cos(\Delta m^2 A_{12})+   \frac{1}{2}\cos(\Delta b^2 B_{11}) + \frac{1}{2}\cos(\Delta b^2 B_{22}) \Big)\,,
\eeqa
and
\be \label{P_lensing_3_norm}
|N|^2 =\frac{1}{2 \left(1+  \cos^2\alpha \cos(\Delta b^2 B_{11})+ \sin^2\alpha \cos(\Delta b^2 B_{22})\right)},
\ee
where $\Delta m^2=\Delta m_{21}^2$. The parameters $A_{11,22}$ and $B_{11,22}$ can be written in terms of the independent parameters $A_{12}$ and $B_{12}$ as
$A_{11,22}=A_{12}\mp X \Delta b^2$ and $B_{11,22}=B_{12}\mp X \Delta m^2$ (-sign for  11 and + for 22 components). Here, $X = \frac{r_S+r_D}{8E_0 r_S r_D}$. The noteworthy features of the derived lensing probability are:
\begin{enumerate}[(1)]
\item ${\cal P}^{\rm lens}_{e \mu}$ does not change under the interchange of $b_1$ and $b_2$. It is therefore an even function of $\Delta b^2$.
\item Under the interchange of $m_1$ and $m_{2}$, the probability does not remain the same unless $\Delta b^2=0$ or $\alpha = \pi/4$, due to  $B_{11,22}$ terms. This is in contrast to two flavour oscillation in the flat spacetime in which interchange of $m_{1}$, $m_2$ implies the same probability. Therefore, Eq. (\ref{P_lensing_3}) is sensitive to the neutrino mass ordering and lead to different results for $\Delta m^2>0$ and $\Delta m^2 < 0$\footnote{Note that this feature is not reflected in oscillation probability obtained for the two flavour case in \cite{Fornengo:1996ef} as the expression derived there does not include appropriate normalization factor.}.
\item The lensing probability, ${\cal P}^{\rm lens}_{e \mu}$, is not only sensitive to the mass ordering but it also explicitly depends on the sum of squared neutrino masses through $B_{12}$ in general.
\end{enumerate}

The above features become more clear when Eq. (\ref{P_lensing_3}) is expanded for small $\Delta b^2$. Defining a dimensionless parameter $\epsilon \equiv \Delta b^2 B_{12}$, and for $\epsilon \ll 1$ Eq. (\ref{P_lensing_3}) can be approximated as
\beqa \label{P_lensing_Approx}
{\cal P}^{\rm lens}_{e \mu} & \approx & \sin^2 2\alpha\, \sin^2\left(\Delta m^2 \frac{ A_{12}}{2} \right)\, \left( 1 - \frac{\epsilon^2}{2} \frac{\Delta m^2}{\sum m^2} \cos2\alpha\, + \frac{\epsilon^4}{16} \frac{\Delta m^2}{\sum m^2} \right.\, \nonumber \\ 
& \times & \left.   \left(\frac{\Delta m^2}{\sum m^2} \left(2 \cos 4\alpha + \csc^2 \left(\Delta m^2\frac{A_{12}}{2}\right) \right)- \frac{2}{3} \left(1+\left(\frac{\Delta m^2}{\sum m^2}\right)^2 \right) \cos 2\alpha \right)\,+ {\cal O}(\epsilon^6) \right)\,,\eeqa
where $\sum m^2 = m_1^2+m_2^2$. Thus, we see that the probability of transition has a clear dependency on $\sum m^2$ as well, unless one adopts certain very special trajectories which only depend on $\Delta m^2$ (these are non-geodetic in nature generically \footnote{It is easy to verify that Keplerian orbits $ r (\phi) = a/(1 + e \cos{(\phi-\phi_0)})$ maintain the dependency on $\sum m^2$. Therefore, a detector (such as on earth in its orbit around the sun) moving along its geodesic would be sensitive to the individual masses of neutrino through neutrino oscillations.}). Furthermore, as mentioned earlier, ${\cal P}^{\rm lens}_{e \mu}$ is sensitive to the sign of $\Delta m^2$ if the mixing angle  is not maximal, $\alpha \neq \pi/4$.  For the maximal case ($\alpha =\pi/4$), the oscillation probability Eq. (\ref{P_lensing_Approx}) does not depend on the $\sum m^2$ at ${\cal O}(\epsilon^2)$ and the mass dependent lensing effects arise through the higher order terms in $\epsilon$. In fact, the absence of signature dependency or order $\epsilon^2$ effects, for 
maximal case can be attributed to the normalization Eq. (\ref{Normalization}) which contributes also in $\epsilon^2$ order. 

The role of normalization has also an observable effect, as with the correct normalization,  the total probability of transition and survival of an initial flavor species will add to unity. Generically, the ratio of survival and transition probability, which depends on $\sum m^2$ and keeps oscillating over the trajectory, provides a natural measurement handle for the masses of the neutrino. As an example, the {\it flipping points}, specified by ($r_D, \phi_D^e$)  on the trajectory where the two probabilities become equal, provide the information of $\sum m^2$ (see Appendix \ref{App:Flipping}). For $\alpha \neq \pi/4$, we have
\begin{widetext}
\bea
\sum m^2 = \frac{16 n \pi E_0 r_S r_D}{Z \Delta b^2}  +  \frac{16 E_0 r_S r_D}{Z \Delta b^2}\cot^{-1}\left(\tan\left(\frac{\Delta m^2 Z \Delta b^2 }{16 E_0 r_S r_D}\right) G(r_S, r_D,  \phi_D^e,\alpha, \Delta m^2, R_s) \right), \nonumber\\ \label{Flip1M}
\eea
where  $G(r_S, r_D,  \phi_D^e,\alpha, \Delta m^2, R_s)$ can take two values $G_+$ or $G_-$ with 
\bea
G_{\pm}(r_S, r_D,  \phi_D^e,\alpha, \Delta m^2, R_s) &\equiv& \frac{2 \cos2\alpha \cos^2\alpha-\sin^2 2\alpha\cos\zeta \pm \sqrt{\sin^4 2\alpha\cos^2\zeta+\sin^2 2\alpha \cos^2 2\alpha}}{ 2\cos2\alpha \cos^2\alpha+\sin^2 2\alpha\cos\zeta \mp \sqrt{\sin^4 2\alpha\cos^2\zeta+\sin^2 2\alpha \cos^2 2\alpha}}, \nonumber\\
\eea
\end{widetext}
where $+$ and $-$ symbols appearing in the expression above through $\pm$ belong to $G_+$ and $G_-$ respectively. Further,  $\zeta \equiv \Delta m^{2}\left(r_S + r_D+R_{s}- (r_S + r_D) \sum b^{2}/4 r_{S} r_{D}\right)/2 E_0 $ , $Z = r_S +r_D$ and $n \in \mathbb{Z}$. Whereas, for $\alpha =\pi/4$, we simply get
\begin{equation} 
\sum m^2 = \frac{(2n+1)8   E_0 r_S r_D}{Z \Delta b^2}\pi \pm \Delta m^2. \label{Flip2M}
\end{equation}
Clearly, the right hand sides of the expressions Eq. (\ref{Flip1M}, \ref{Flip2M}) for $\sum m^2$ depend completely on observationally deterministic quantities and hence the identification of flipping point, therefore, gives its value off. Provided with $\Delta m^2$ one thus obtains the individual masses of both the species.

For a quantitative understanding of the mass dependent effects in Eq. (\ref{P_lensing_3}), it is useful to obtain the impact parameter in terms of the geometrical parameters of the system. A detailed description of lensing phenomena  is described in Fig. \ref{fig:lensing01}. 
Consider $(x',y')$ coordinate system obtained by rotating original $(x,y)$ coordinates by angle $\varphi$ such that $x'=x \cos\varphi + y \sin\varphi$ and $y'=-x \sin\varphi + y \cos\varphi$. In the rotated frame, the angle of deflection of neutrino from its original path with impact parameter $b$ is obtained as 
\be
\delta \sim \frac{y'_D-b}{x'_D}= -\frac{2 R_s}{b},
\ee
where $(x'_D,y'_D)$ is location of the detector. In the second equality we use the expression for $\delta$ in weak lensing ($|b| \gg R_s$) limit. Using the identity $\sin\varphi = b/r_S$, we obtain
\be \label{b_eq}
\left(2 R_s x_D + b y_D \right) \sqrt{1-\frac{b^2}{r_S^2}} = b^2 \left(\frac{x_D}{r_S} +1 \right)-\frac{ 2 R_s b y_D}{r_S}\,.
\ee
Solutions of the above equation give the impact parameters in terms of the geometrical parameters such as $r_S$, $R_s$ and the lensing location $(x_D,y_D)$. In equatorial plane $(\theta=\pi/2)$ and for $y_D \ll x_D$, the solutions of the above equation imply 
\begin{equation}\label{eq:34}
\sum b^2 \approx 4R_s x_D \left(1+\frac{y_D^2}{4R_s x_D}\right)\,,~~~\Delta b^2 \approx -y_D\sqrt{8R_s x_D}\,\left(1+\frac{y_D^2}{16R_s x_D}\right)\,.
\end{equation}

\begin{figure}[t!]
\centering
\subfigure{\includegraphics[width=0.95\textwidth]{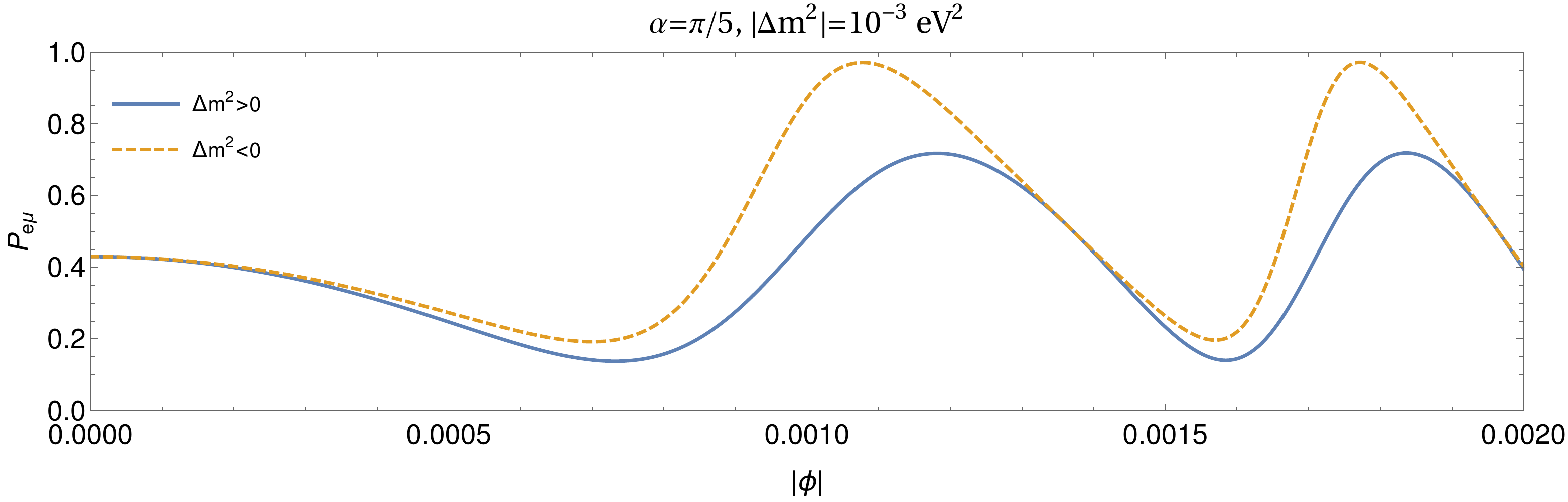}}
\subfigure{\includegraphics[width=0.95\textwidth]{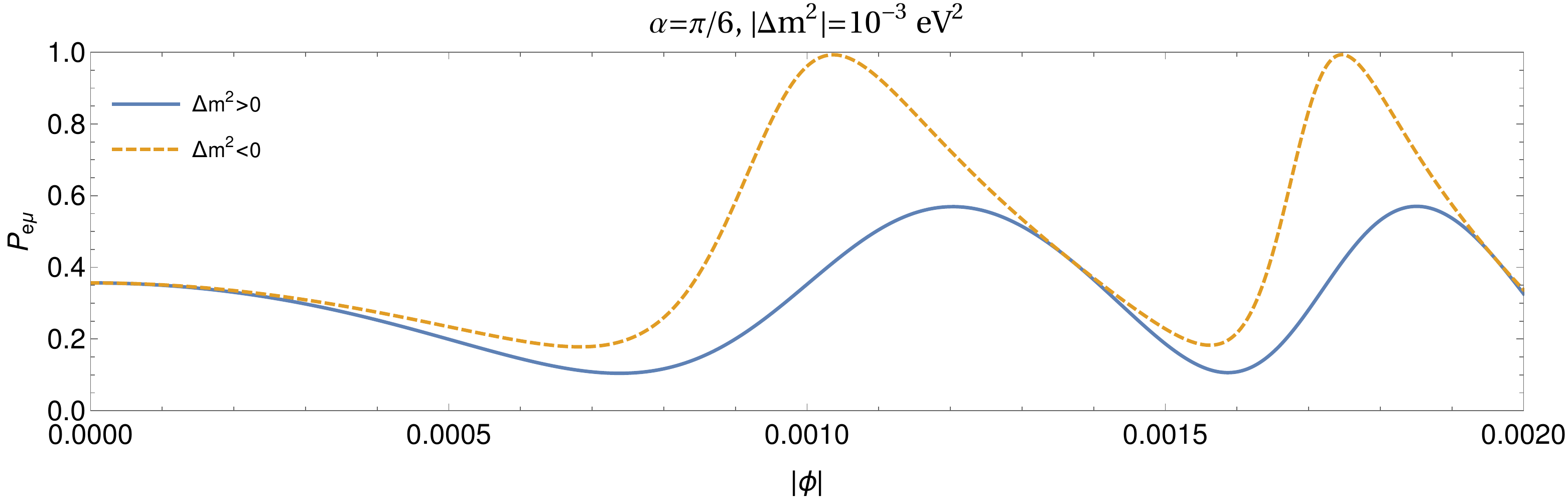}}
\caption{Probability of $\nu_e \to \nu_\mu$ conversion as function of azimuthal angle $\phi$ for the normal and inverted ordering of neutrino masses in the two flavour case. Here, $R_s = 3$ km, $r_D=10^8$ km, $r_S=10^5 r_D$ and $E_0 = 10$ MeV. The lightest neutrino is assumed to be massless.}
\label{fig:2F_ordering}
\end{figure}

\begin{figure}[t!]
\centering
\subfigure{\includegraphics[width=0.95\textwidth]{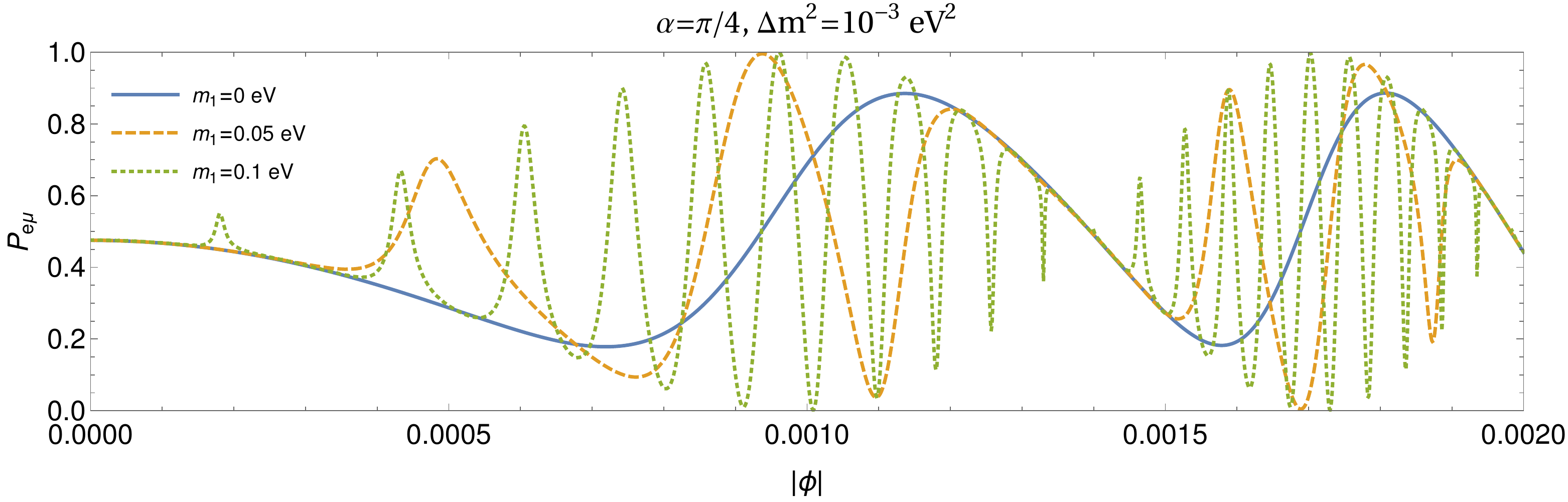}}
\subfigure{\includegraphics[width=0.95\textwidth]{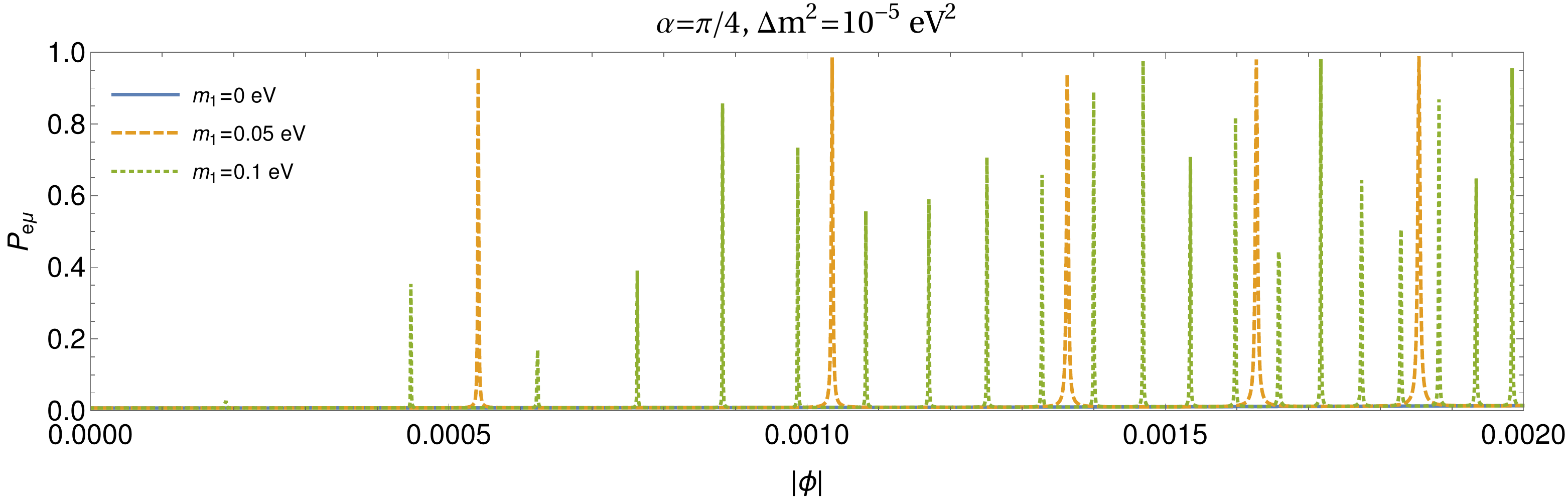}}
\subfigure{\includegraphics[width=0.95\textwidth]{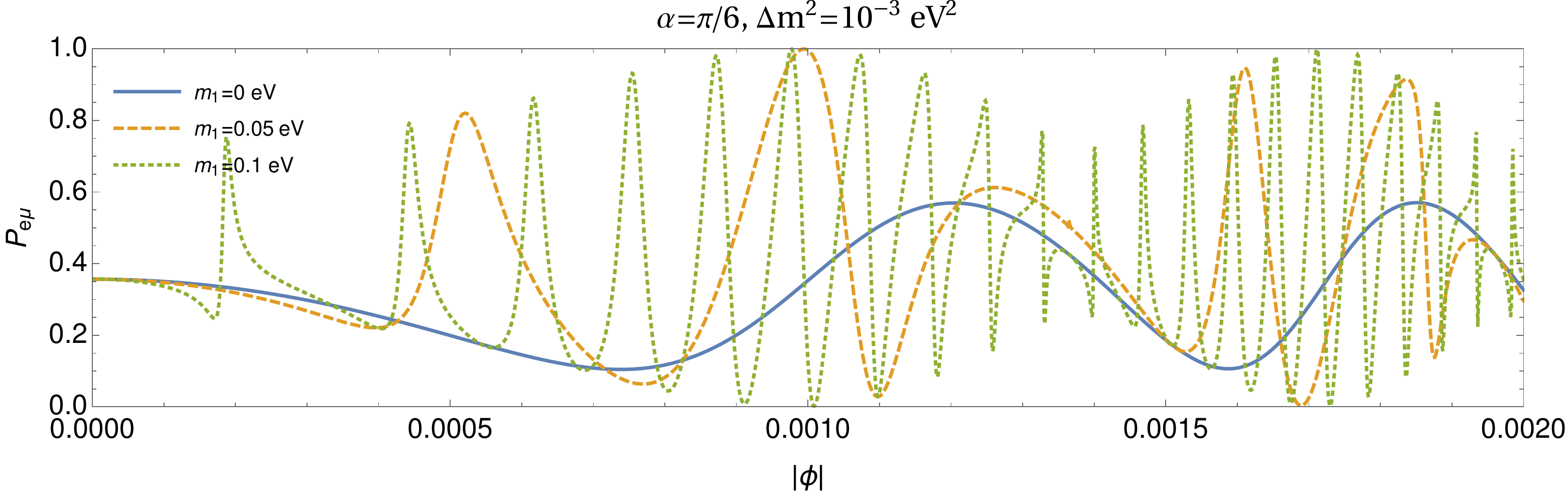}}
\caption{Probability of $\nu_e \to \nu_\mu$ conversion as function of azimuthal angle $\phi$ for different different values of $m_1$ and for $\Delta m^2>0$. Here, $R_s = 3$ km, $r_D=10^8$ km, $r_S/r_D=10^5$ and $E_0 = 10$ MeV.}
\label{fig:2F_mass}
\end{figure}

\begin{figure}[t!]
\centering
\subfigure{\includegraphics[width=0.99\textwidth]{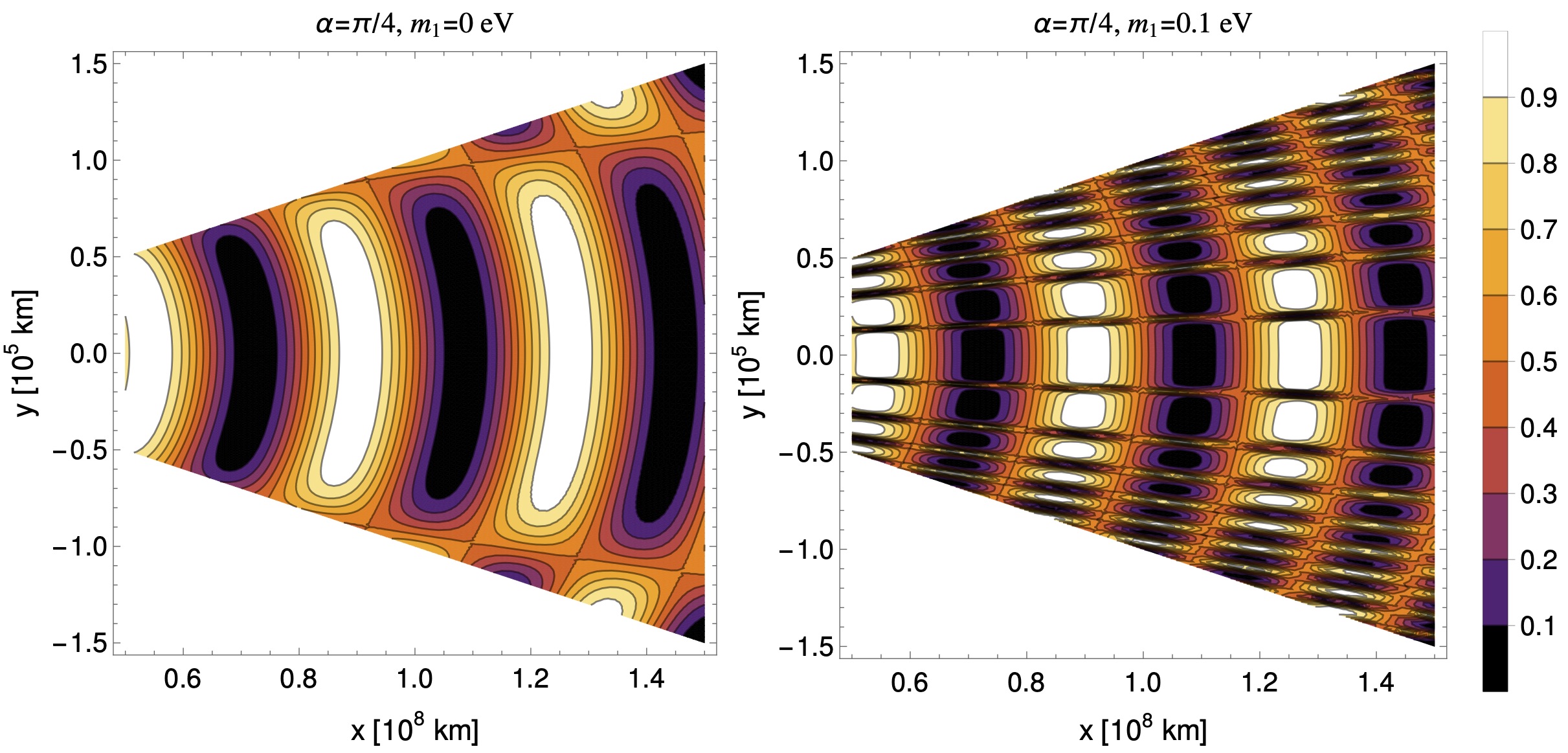}}
\caption{Contours of $\nu_e \to \nu_\mu$ conversion probability in the presence of gravitational lensing for $R_s = 3$ km, $r_D=10^8$ km, $r_S=10^5\, r_D$ and $E_0 = 10$ MeV. Here, $(x_D,y_D)$ is location of observer in equatorial plane.}
\label{fig:2F_3D}
\end{figure}

We now analyse the two flavour case by evaluating the oscillation probability given in Eq. (\ref{P_lensing_3}) through solving Eq. (\ref{b_eq}) numerically in the equatorial plane. The values of geometrical parameters are chosen to simulate the Sun-Earth system in which the Sun is taken as the gravitational lens while the Earth represents the location of the detector. We assume a circular trajectory for the Earth around the Sun (with $x_D = r_D \cos\phi$, $y_D=r_D \sin \phi$) and take $r_D=10^8$ km, $R_s = 3$ km. The source of neutrinos is assumed to be located at $r_S = 10^5\, r_D$ on the opposite side of the Sun and it emits relativistic neutrinos with common $E_0 =10$ MeV. In our analysis, we compute ${\cal P}^{\rm lens}_{e \mu}$ only for those values of $b_{1,2}$ which justify the approximation, $R_s \ll b_{1,2} \ll r_D$, used while deriving Eq. (\ref{Phi_non-radial}) and Eq. (\ref{b_eq}). The results are displayed in Figs. \ref{fig:2F_ordering}, \ref{fig:2F_mass} and \ref{fig:2F_3D}\footnote{ For prediction in a realistic settings, one will also need to account for the matter interaction effects once the neutrino passes through the sun, since $b<R_{sun}$. Such matter interaction inside the sun is already studied, see e.g. \cite{Smirnov:2004zv}.  For realizing pure Schwarzschild solution discussed here, one needs to move slightly farther on the x-axis i.e. increase $x_D$. }.  In these figures, we explore the neutrino flavour conversion over azimuthal angle $\phi$ over the range of 0.002 radians, which corresponds to roughly 3 hours into the trajectory of the earth around the sun.

The dependence of conversion probability on the neutrino mass ordering for non-maximal $\alpha$ is shown in Fig. \ref{fig:2F_ordering}. Away from $\phi=0$, the observer is not in the same line of the source and gravitational object which give rise to different probabilities for normal and inverted mass orderings. For $\alpha < \pi/4$, the inverted ordering always corresponds to relatively increased conversion probability than that of the normal ordering. This is due to the fact that $B_{11} < B_{22}$ in Eq. (\ref{P_lensing_3}) for inverted ordering enhances ${\cal P}^{\rm lens}_{e \mu}$ for $\alpha < \pi/4$. This trend gets reversed if $\alpha > \pi/4$ as Eq. (\ref{P_lensing_3}) is invariant under simultaneous transformations, $m_{1,2} \to m_{2,1}$ and $\alpha \to \pi/2-\alpha$. It is clear that if the mixing angle is different from $\pi/4$, one can infer the neutrino mass ordering from the lensing effects even in two flavour case in contrast to the standard neutrino oscillations in vacuum 
without lensing.  On the paths independent of $\sum m^2$, e.g. along $\phi =0$, the probabilities depend upon (and thus reveal) the standard neutrino oscillation parameters (path length, $\Delta m^2$, energy). However, as discussed above, there are other paths which carry imprint of absolute masses as well as the signature of $\Delta m^2$ depending upon the mixing angle. Thus, overall observation of probabilities along various directions in the plane of lensing is quite resourceful for neutrino physics.

The lensing of neutrinos is not only sensitive to the neutrino mass ordering but it can also shed light on the absolute mass scale of the neutrinos.  As it is shown in Fig. \ref{fig:2F_mass}, for fixed $\alpha$ and $\Delta m^2$, the probability itself oscillates as one goes further from $\phi=0$. The frequency of these oscillations depend on the absolute mass scale of the neutrino. We find that the probability oscillates slowly for hierarchical neutrinos, i.e. $m_1 \ll \sqrt{\Delta  m^2}$, when compared to the case when neutrinos are almost degenerate, i.e. $m_1 \gsim \sqrt{\Delta  m^2}$. This qualitative feature is more or less independent from the specific values of $\alpha$ and $\Delta m^2$ as it can be seen from Fig. \ref{fig:2F_mass}. In the two flavour case, it is therefore possible to infer the absolute neutrino masses by measuring the neutrino transition or survival probabilities with respect to the angle $\phi$. 

We also show the probability distribution as function of $(x_D,y_D)$  in equatorial plane in Fig. \ref{fig:2F_3D}. The values of $y_D$ and $x_D$ are chosen such that they satisfy the weak lensing limit. As it is expected, the maxima and minima of the probability along $y_D=0$ line do not depend on the  absolute value of neutrino masses. The lensing effect however give rises to different frequencies of maxima and minima of oscillation probabilities for different values of absolute neutrino mass as one deviates from $y_D=0$.

\section{Three flavour case: Numerical Results}
\label{sec:3FNum}
\begin{figure}[!ht]
\centering
\subfigure{\includegraphics[width=0.95\textwidth]{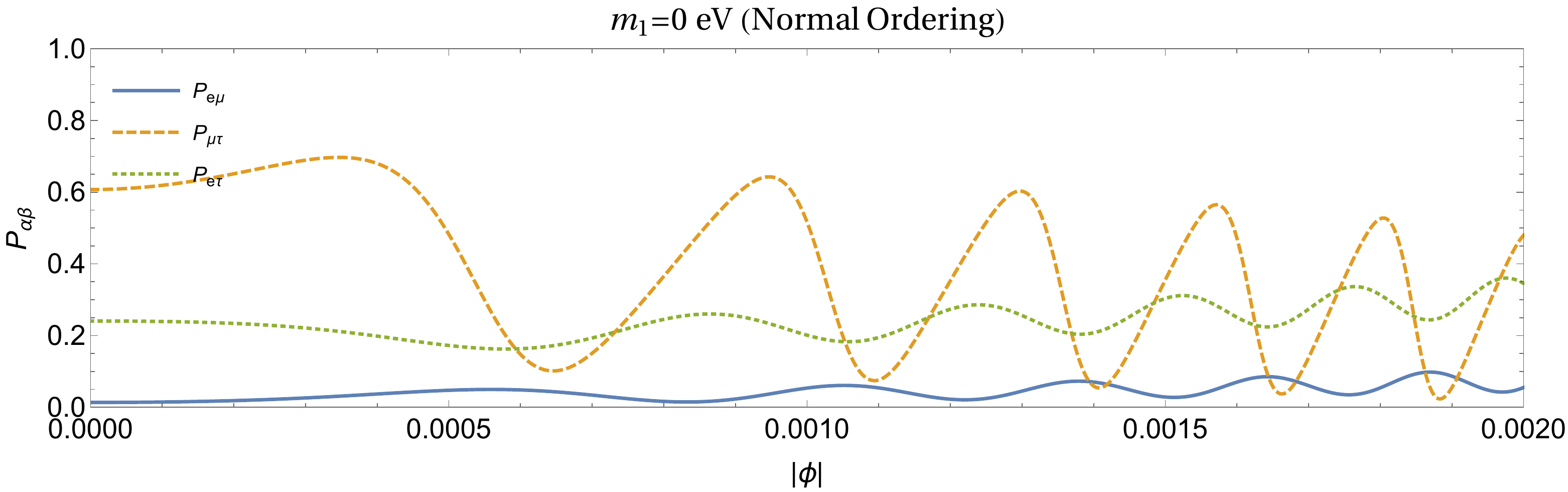}}
\subfigure{\includegraphics[width=0.95\textwidth]{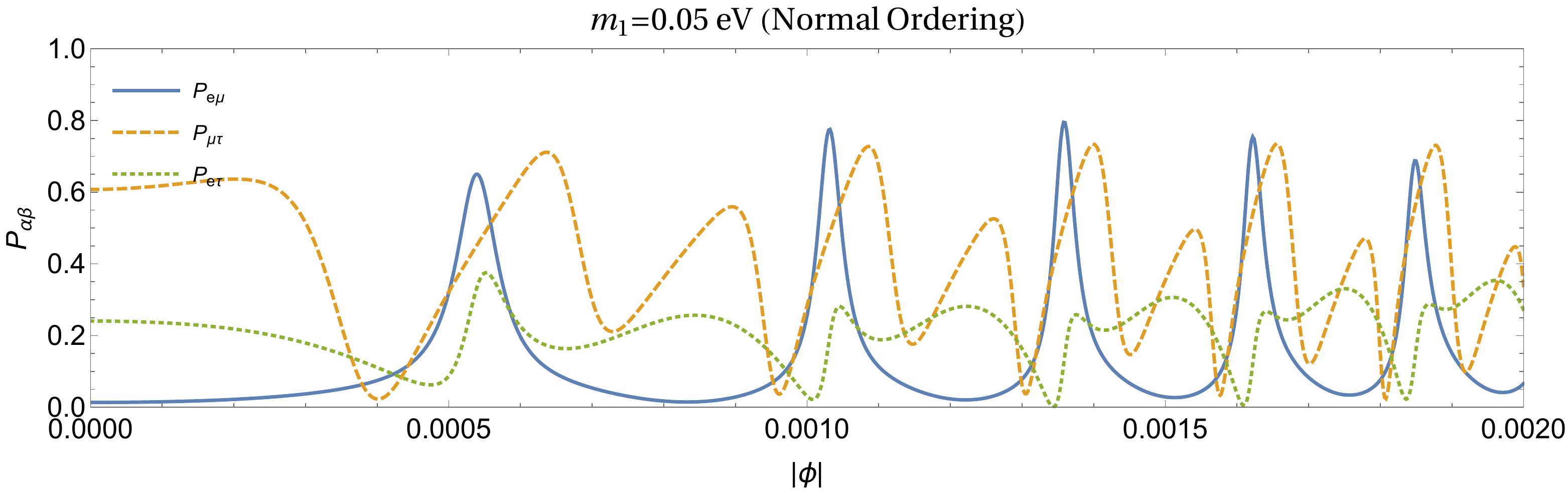}}
\subfigure{\includegraphics[width=0.95\textwidth]{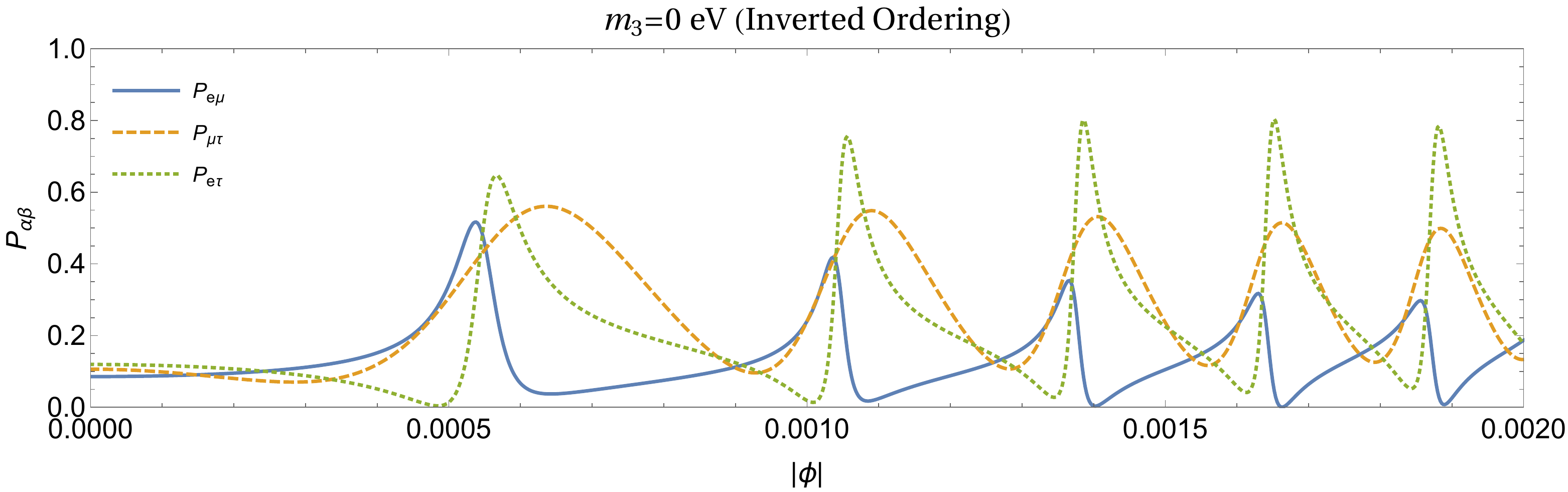}}
\subfigure{\includegraphics[width=0.95\textwidth]{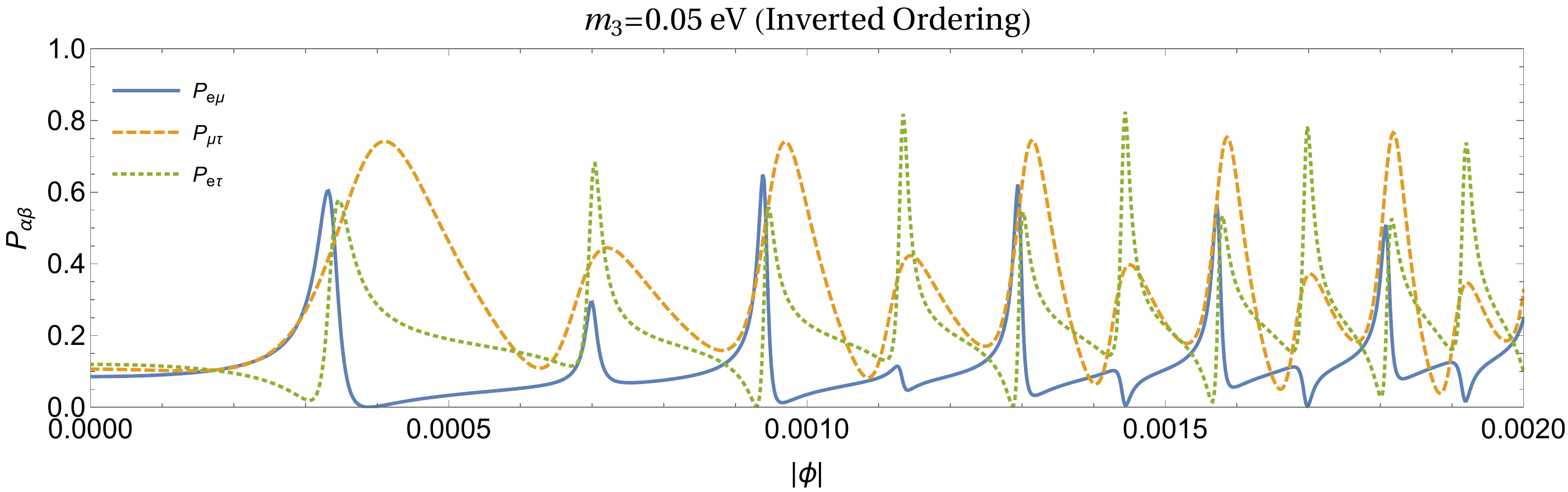}}
\caption{Probability of $\nu_\alpha \to \nu_\beta$ conversion, for different $\alpha$ and $\beta$, as function of azimuthal angle $\phi$ in three flavour case. We take $R_s = 3$ km, $r_D=10^8$ km, $r_S=10^5 r_D$ and $E_0 = 10$ MeV. The values of neutrino mass squared differences, mixing angle and the Dirac CP phase are taken from the latest (NuFIT 4.1 (2019)) global fit \cite{Esteban:2018azc}.}
\label{fig:3F_mass}
\end{figure}

In this section, we report results obtained for neutrino lensing in the three flavour case. The geometrical setup and the values of parameters used are same as in the two flavour case discussed in the previous section. For neutrino masses and mixing parameters, we use the values from the latest fit (NuFIT 4.1 (2019)) of neutrino oscillation data \cite{Esteban:2018azc}. These are $\Delta m_{21}^2 = 7.39 \times 10^{-5}~{\rm eV}^2$,  $\theta_{12}=33.82^{\circ}$, $\Delta m_{31}^2 = 2.523 \times 10^{-3}~{\rm eV}^2$ ($\Delta m_{32}^2 = - 2.510 \times 10^{-3}~{\rm eV}^2$), $\theta_{23}=48.3^{\circ}$ ($\theta_{23}=48.8^{\circ}$), $\theta_{13}=8.61^{\circ}$ ($\theta_{13}=8.64^{\circ}$), $\delta_{\rm CP}=222^{\circ}$ ($\delta_{\rm CP}=282^{\circ}$) for normal (inverted) ordering. The mixing matrix $U$ in three flavour case is evaluated using these mixing parameters and used to obtain the oscillation probabilities using Eq. (\ref{P_lensing_2}) in equatorial plane. We compute conversion probabilities for 
$\nu_e \to \nu_\mu$, $\nu_\mu \to \nu_\tau$ and $\nu_e \to \nu_\tau$ for normal and inverted ordering and for two values of the lightest neutrino masses in each case. The results are shown in Fig. \ref{fig:3F_mass}.

It can be noticed that the frequency of the oscillations of probabilities increases as the lightest neutrino mass is increased from 0 to 0.05 eV in both the cases of normal and inverted ordering. This is qualitatively in agreement with the two flavour case. Unlike in the two flavour case, one obtains different values of transition probabilities for normal and inverted ordering even for $\phi=0$. Different ${\cal P}_{\alpha \beta}$ then oscillate differently for nonzero $\phi$ as it can be seen from Fig. \ref{fig:3F_mass}. The mass of the lightest neutrino in both the cases can be inferred by measuring the frequency of the oscillation of transition probabilities.

Throughout this paper, the neutrino wave functions are assumed plane waves. A more realistic treatment of the neutrino oscillations would involve wave packet approach in which neutrinos are produced and detected as wave packets of finite sizes in the position space \cite{Giunti:1997wq,Akhmedov:2009rb,Akhmedov:2010ua}. This can give rise to decoherence of neutrinos when propagation over long distances is involved which ultimately leads to wash-out of the flavour oscillations. In flat spacetime and for the gaussian wave packets, the length over which neutrinos maintain coherence is typically given by 
\be \label{LCoh_flat}L_{\rm coh} \simeq  4 \sqrt{2} \sigma_x E_0^2/|\Delta m_{ij}^2|\,, \ee
where, $\sigma_x^2 \equiv \sigma_{xS}^2 + \sigma_{xD}^2$. Here, $\sigma_{xS}$ represents the position space width of the neutrino wave packet produced at the source while $\sigma_{xD}$ denotes the same for the neutrino at the detector \cite{Giunti:1997wq}. The same expression of coherence length continues to hold in the case of Schwarzschild geometry but with the flat spacetime distance $L_{\rm coh}$ replaced by coordinate distance \cite{Chatelain:2019nkf}.

Following these results we find that the coherence condition, in the weak gravity limit, is given by
\be \label{coherence_cond}
(r_S + r_D) \left(1- \frac{b^2}{2 r_S r_D} + \frac{2 GM}{r_S + r_D} \right)\,  <\,  \frac{4 \sqrt{2} \sigma_x E_0^2}{|\Delta m_{ij}^2|}\,.
\ee
Further, the terms of ${\cal O}\left(\frac{b^2}{r_S r_D}\right)$ and ${\cal O}\left(\frac{GM}{r_S + r_D}\right)$ in the above are negligible for the weak lensing cases. Therefore, the treatment of decoherence of neutrinos in the presence of weak lensing is similar to that in flat spacetime. For the typical values of parameters considered in the paper, i.e. $r_S + r_D \simeq 10^{13}$ km, $E_0 \simeq 10$ MeV and $|\Delta m_{ij}^2| \simeq 10^{-3}$ eV$^2$, one obtains $\sigma_x \simeq 1.8$ cm. Using $\sigma_x \sigma_p \sim \hbar$ where $\sigma_p$ is the width of neutrino wave packet in the momentum space, the above value of $\sigma_x$ would imply neutrino wave packets with $\sigma_p/p \lesssim {\cal O}(10^{-11})$ at the source and detector in order to satisfy the coherence condition, Eq. (\ref{coherence_cond}). Therefore, this would require very precise information about the energies or meomenta of particles involved in the production and detection processes of neutrinos.

\section{Conclusions}
\label{sec:concl}
Neutrino flavour oscillation in flat spacetime is known to depend on the difference between the squared masses and not on the individual masses of neutrinos. We show that weak gravitational  lensing of neutrinos modifies this standard picture drastically. The oscillation probabilities evaluated in the presence of lensing introduces novel effects which depend on the absolute neutrino mass scale in general. We demonstrate this explicitly considering a Schwarzschild mass as a gravitational source for lensing, while the source and detector for neutrinos are kept at finite distances from it. In the presence of Schwarzschild mass, the neutrinos produced at source take more than one classical path to reach to the detector through weak lensing, as justified by the Eikonal approximation. Neutrinos travelling along different trajectories develop different phases which give rise to interference at the location of the detector. We show that the phase difference not only depends on difference of squared neutrino masses but also depends on the individual neutrino masses in general. The dependency on the individual neutrino masses survives for wide class of trajectories of the detector including the geodesic ones. Therefore, the detectors situated on the Keplerian orbits around the Schwarzschild mass will be capable of revealing the absolute mass of neutrinos through measurements of flavour transition probabilities.

We derive the general expression of interference of oscillation probability, valid in the weak lensing limit, for $N$ neutrino flavours and arbitrary number of paths. We study the interference pattern in detail considering the equatorial plane for simplicity in which neutrinos are confined to travel along two trajectories. For $N=2$, we show that the lensing probability is sensitive to the sign of $\Delta m^2$ unlike in the case of standard neutrino oscillation in vacuum. Further, it also depends on the individual masses when source, gravitating object and detector are not collinear. flavour transition or survival probabilities oscillate as a result of interference when the detector moves away from the collinear axis. The frequency of these oscillations of probabilities depend on the absolute mass scale of neutrinos. We find that the hierarchical neutrinos, i.e. $m_1 \ll m_2 $, give rise to slower oscillations of probability in comparison to the case when they are almost degenerate, i.e. $m_1 \simeq m_2$. We 
study these effects quantitatively considering the sun-earth like system in which the sun plays role of Schwarzschild mass for neutrinos coming from a distant source. We also give numerical results for $N=3$ which also captures the qualitative effects of lensing obtained in the case of two flavours.  All these novel effects are indeed gravitationally induced as they vanish in the flat space ($R_s \to 0$)  limit. We, therefore conclude that the study of effects of gravitation in the standard analysis of neutrino oscillations can be very resourceful and informative.

Although our results reveal some very non-trivial aspects of neutrino oscillation in the presence of gravitational lensing, a more careful realistic treatment would be required before they can be used for real experimental tests. Firstly, we have adopted the standard practice of using quantum mechanical treatment for the neutrinos in the present study. However, justifiably a quantum field theoretic treatment \cite{Grimus:1998uh, Kobach:2017osm,Grimus:2019hlq} would be more appropriate. Such a treatment would  take into account the mode propagation and will take us out of the limitation of Eikonal validity of using classical trajectories. It may also provide insight to novel phenomenon of gravitational particle creation and the resulting mixing of flavour as well as energy modes due to this particle creation \cite{Blasone:2017nbf, Blasone:2018iih, Cozzella:2018qew, Blasone:2018czm, Blasone:2019agu}, etc.. A more realistic treatment of neutrino oscillation with/without lensing should consider the wave packet approach. This, however, relies on the details of the exact mechanisms of neutrino production and detection. As mentioned earlier, such a treatment naturally accounts for the decoherence effects and washing out of oscillation.  A detailed study addressing all the above aspects is beyond the scope of this paper and it should be taken up elsewhere.

\acknowledgments
The authors thank Subhendra Mohanty and T. Padmanabhan for careful readings of the manuscript and for useful comments. We also thank Pratibha Jangra for helpful discussions. The authors thank Hrishikesh Chakrabarty for pointing out an error in one of the expressions in the previous version. HS would like to thank Council of Scientific \& Industrial Research (CSIR), India for the financial support through research fellowship Award No. 09/947(0081)/2017-EMR-1. Research of KL is partially supported by the Department of Science and Technology (DST) of the Government of India through a research grant under INSPIRE Faculty Award (DST/INSPIRE/04/2016/000571). KMP is partially supported by a research grant under INSPIRE Faculty Award (DST/INSPIRE/04/2015/000508) from the DST, Government of India. HS and KL are also grateful towards the hospitality of Physical Research Laboratory,  Ahmedabad, where part of this work was carried out.

\appendix
\section{Eikonal Approximation}
\label{App:Eikonal}
The  classical Hamilton Jacobi function $S(x,p) = \int p_{\mu} dx^{\mu}$ satisfies
$\partial S/ \partial x^{\mu} = p_{\mu}$.
Therefore, classically we have,
\be
p_{\mu}p^{\mu}=m^2 \Rightarrow g^{\mu \nu} \frac{\partial S}{\partial x^{\mu}}\frac{\partial S}{\partial x^{\nu}} -m^2 =0.
\ee
On the other hand the squared Dirac equation satisfies,
\be
\left[\partial_{\mu} \partial^{\mu} +m^2\right]\psi =0.
\ee
If the wavefunction is taken as $\psi = e^{i S/\hbar}$ then
\be
\left[g^{\mu \nu} \frac{\partial S}{\partial x^{\mu}}\frac{\partial S}{\partial x^{\nu}}-i \hbar g^{\mu \nu} \partial_{\mu} \partial_{\nu}S -m^2\right]\psi =0.
\ee
Therefore, as long as 
$$\frac{\hbar g^{\mu \nu} \partial_{\mu} \partial_{\nu}S}{ g^{\mu \nu} \partial_{\mu} S \partial_{\nu}S}\ll 1, $$
the classical $S$  is a good approximation for the phase of the neutrino. For the massless case $S$ is identically zero and therefore one needs to evaluate the ratio in the limit of small mass $m$. Since for a massive particle
\be
S= \Phi(r) =\frac{m^2}{2 E}\left[ r_S + r - \frac{b^2 (r_S + r)}{2r_S r} + 2 GM \right],
\ee
a naive estimate suggests that the Eikonal approximation will be valid as long as at all the points ($r$) along the path, the condition
\be
\left( 1 -\frac{2 GM}{r} \right)\frac{\hbar }{2E}\frac{b^2}{ r^3}\ll 1,
\ee
is satisfied. This becomes increasingly valid in weak lensing limit.

\section{Observable sensitive to the individual masses: Angle of flavour flipping}\label{App:Flipping}
An interesting aspect of the transition probability is that in two flavour case Eq. (\ref{P_lensing_3}), there exists a point where the transition probability from species $\alpha$ to $\beta$ becomes equal to its survival probability $\alpha \rightarrow \alpha $. In other words, one of these two processes remain dominant over some portion of trajectories (indicated by angle $\phi$ in the geodesic trajectory of the earth around the sun in sun-earth system) and then become sub dominant for some other portion. In principle this flipping over occurs multiple times in a trajectory. Therefore, this angle of flipping over (an observable) also carries an imprint of $\sum m^2$ which we can utilize and find out the absolute masses. Flipping occurs at point when  probability transition and survival probability become equal. For two flavour case, the survival probability is obtained as,
\bea
 P_{\alpha \rightarrow \alpha} = 4N^2\Bigg(\cos^4 \alpha \cos^2 \left(m_1^2\frac{Z\Delta b^2}{8 E_0 r_S r_D}\right) + \sin^4 \alpha \cos^2 \left(m_2^2\frac{Z\Delta b^2}{8 E_0 r_S r_D}\right) \nonumber\\
	 + 2\sin^2\alpha \cos^2\alpha  \cos \left(m_1^2\frac{Z\Delta b^2}{8E_0r_S r_D}\right) \cos \left(m_2^2\frac{Z\Delta b^2}{8E_0r_S r_D}\right)\cos\zeta\Bigg),
\eea
and the transition probability expression (\ref{P_lensing_3}) can be cast as
\bea 
P_{\alpha \rightarrow \beta} = 4N^2 \sin^2\alpha \cos^2\alpha \Bigg(\cos^2 \left(m_1^2\frac{Z\Delta b^2}{8E_0r_S r_D}\right) + \cos^2 \left(m_2^2\frac{Z\Delta b^2}{8E_0r_S r_D}\right)	\nonumber\\
-2 \cos \left(m_1^2\frac{Z\Delta b^2}{8E_0r_S r_D}\right) \cos \left(m_2^2\frac{Z\Delta b^2}{8E_0 r_S r_D}\right)\cos\zeta\Bigg),
\eea
where $\zeta \equiv \frac{\Delta m^{2}}{2 E_0}\left(Z+R_{s}-\frac{Z \sum b^{2}}{4 r_{S} r_{D}}\right) $ and $Z= r_S + r_D$.
Therefore, the condition for flipping over is obtained as 
\begin{equation} \label{Eq. 10}
 \cos2\alpha \cos^2\alpha \frac{\cos \left(m_1^2\frac{Z\Delta b^2}{8E_0r_S r_D}\right)}{\cos \left(m_2^2\frac{Z\Delta b^2}{8E_0r_S r_D}\right)} -\cos2\alpha \sin^2\alpha \frac{\cos \left(m_2^2\frac{Z\Delta b^2}{8E_0r_S r_D}\right)}{\cos \left(m_1^2\frac{Z\Delta b^2}{8E_0r_S r_D}\right)}  + \sin^22\alpha \cos\zeta =0.
\end{equation}
For $\alpha \neq \pi/4$, the above equation can be inverted as
\bea
\sum m^2 = \frac{16 n \pi E_0 r_S r_D}{Z \Delta b^2}  +  \frac{16 E_0 r_S r_D}{Z \Delta b^2}\cot^{-1}\left(\tan\left(\frac{\Delta m^2 Z \Delta b^2 }{16 E_0 r_S r_D}\right) G(r_S, r_D, \phi,\alpha, \Delta m^2, R_s) \right), \nonumber\\\label{Flip1}
\eea
where 
\bea
G(r_S, r_D, \phi,\alpha, \Delta m^2, R_s) &\equiv& \frac{2 \cos2\alpha \cos^2\alpha-\sin^2 2\alpha\cos\zeta \pm \sqrt{\sin^4 2\alpha\cos^2\zeta+\sin^2 2\alpha \cos^2 2\alpha}}{ 2\cos2\alpha \cos^2\alpha+\sin^2 2\alpha\cos\zeta \mp \sqrt{\sin^4 2\alpha\cos^2\zeta+\sin^2 2\alpha \cos^2 2\alpha}}. \nonumber\\
\eea

\noindent Similarly,  for $\alpha = \pi/4$ we get 
\begin{equation} \label{Eq. 25}
\left(\cos\left(\frac{\sum m^2 Z \Delta b^2 }{8 E_0 r_S r_D}\right) + \cos\left(\frac{\Delta m^2 Z \Delta b^2 }{8 E_0 r_S r_D}\right)\right)\cos\zeta = 0.
\end{equation}
Flipping point is given by above equation when either first term or second term is zero, or both.  
If we find point where first term if zero then, we have
\begin{equation} 
\sum m^2 = \frac{(2n+1)8   E_0 r_S r_D}{Z \Delta b^2}\pi \pm \Delta m^2, \label{Flip2}
\end{equation}
since the cosine term is symmetric under $\Delta m^2 \to - \Delta m^2$.  Therefore,
\begin{equation} \label{Eq. 27}
m_{1,2}^2 = \left(\frac{(2n +1)8E_0 r_S r_D}{Z \Delta b^2} \right)\frac{\pi}{2}.
\end{equation}
For $\cos\zeta =0,$ we cannot determine value of $\sum m^2$ because $\zeta $ only involves $\Delta m^2, E_0$ and other terms  which involves geometrical points and geometry.
As can be seen, in \eq{Flip1} and \eq{Flip2} all the direct observable quantities are on the right hand sides, while  $n \in \mathcal{Z}$. Thus, the observation of flip angle $\phi$ for known $r_S, r_D, \alpha, E_0, R_s \text{ and } \Delta m^2$ will reveal $\sum m^2$ (upto an ambiguity due to undetermined value of $n$, which have to settled through some independent observations such as \cite{Choudhury:2018byy}) which along with information of $\Delta m^2$ can give the absolute masses of the neutrinos.


%

\end{document}